\documentclass[conference]{IEEEtran}

\IEEEoverridecommandlockouts
\usepackage{cite}
\usepackage{amsmath,amssymb,amsfonts}
\usepackage{algorithmic}
\usepackage{graphicx}
\usepackage{textcomp}
\usepackage{xcolor}

\DeclareMathAlphabet{\mathsfbr}{OT1}{cmss}{m}{n}
\SetMathAlphabet{\mathsfbr}{bold}{OT1}{cmss}{bx}{n}
\DeclareRobustCommand{\msf}[1]{%
  \ifcat\noexpand#1\relax\msfgreek{#1}\else\mathsfbr{#1}\fi
}

\makeatletter
\newcommand{\msfgreek}[1]{\csname s\expandafter\@gobble\string#1\endcsname}
\makeatother

\DeclareFontEncoding{LGR}{}{} 
\DeclareSymbolFont{sfgreek}{LGR}{cmss}{m}{n}
\SetSymbolFont{sfgreek}{bold}{LGR}{cmss}{bx}{n}
\DeclareMathSymbol{\salpha}{\mathord}{sfgreek}{`a}
\DeclareMathSymbol{\sbeta}{\mathord}{sfgreek}{`b}
\DeclareMathSymbol{\sgamma}{\mathord}{sfgreek}{`g}
\DeclareMathSymbol{\sdelta}{\mathord}{sfgreek}{`d}
\DeclareMathSymbol{\sepsilon}{\mathord}{sfgreek}{`e}
\DeclareMathSymbol{\szeta}{\mathord}{sfgreek}{`z}
\DeclareMathSymbol{\seta}{\mathord}{sfgreek}{`h}
\DeclareMathSymbol{\stheta}{\mathord}{sfgreek}{`j}
\DeclareMathSymbol{\siota}{\mathord}{sfgreek}{`i}
\DeclareMathSymbol{\skappa}{\mathord}{sfgreek}{`k}
\DeclareMathSymbol{\slambda}{\mathord}{sfgreek}{`l}
\DeclareMathSymbol{\smu}{\mathord}{sfgreek}{`m}
\DeclareMathSymbol{\snu}{\mathord}{sfgreek}{`n}
\DeclareMathSymbol{\sxi}{\mathord}{sfgreek}{`x}
\DeclareMathSymbol{\somicron}{\mathord}{sfgreek}{`o}
\DeclareMathSymbol{\spi}{\mathord}{sfgreek}{`p}
\DeclareMathSymbol{\srho}{\mathord}{sfgreek}{`r}
\DeclareMathSymbol{\ssigma}{\mathord}{sfgreek}{`s}
\DeclareMathSymbol{\stau}{\mathord}{sfgreek}{`t}
\DeclareMathSymbol{\supsilon}{\mathord}{sfgreek}{`u}
\DeclareMathSymbol{\sphi}{\mathord}{sfgreek}{`f}
\DeclareMathSymbol{\schi}{\mathord}{sfgreek}{`q}
\DeclareMathSymbol{\spsi}{\mathord}{sfgreek}{`y}
\DeclareMathSymbol{\somega}{\mathord}{sfgreek}{`w}

\DeclareMathSymbol{\svarsigma}{\mathord}{sfgreek}{`c}

\DeclareMathSymbol{\sGamma}{\mathalpha}{sfgreek}{`G}
\DeclareMathSymbol{\sDelta}{\mathalpha}{sfgreek}{`D}
\DeclareMathSymbol{\sTheta}{\mathalpha}{sfgreek}{`J}
\DeclareMathSymbol{\sLambda}{\mathalpha}{sfgreek}{`L}
\DeclareMathSymbol{\sXi}{\mathalpha}{sfgreek}{`X}
\DeclareMathSymbol{\sPi}{\mathalpha}{sfgreek}{`P}
\DeclareMathSymbol{\sSigma}{\mathalpha}{sfgreek}{`S}
\DeclareMathSymbol{\sUpsilon}{\mathalpha}{sfgreek}{`U}
\DeclareMathSymbol{\sPhi}{\mathalpha}{sfgreek}{`F}
\DeclareMathSymbol{\sPsi}{\mathalpha}{sfgreek}{`Y}
\DeclareMathSymbol{\sOmega}{\mathalpha}{sfgreek}{`W}

\DeclareRobustCommand{\mcal}[1]{%
  \ifcat\noexpand#1\relax\mathnormal{#1}\else\cal{#1}\fi
}
\DeclareRobustCommand{\BM}[1]{%
  \ifcat\noexpand#1\relax\bm{\boldUppercaseItalicGreek{#1}}\else\bm{#1}\fi
}
\makeatletter
\newcommand{\boldUppercaseItalicGreek}[1]{\csname var\expandafter\@gobble\string#1\endcsname}
\makeatother

\newcommand{\V}[1]{\bm{#1}} 
\newcommand{\Set}[1]{{\mcal{#1}}} 

\newcommand{\E}[1]{\mathbb{E}\left\{#1\right\}}

\newcommand{\avg}[1]{\overline{\left\{#1\right\}}}

\usepackage{bm}
\usepackage{color, courier}
\usepackage{psfrag}
\usepackage{acronym}
\usepackage{amsmath}
\usepackage{amssymb}
\usepackage{amsfonts}
\usepackage[colorlinks = true, allcolors = black]{hyperref}
\usepackage{latexsym}
\usepackage{mathrsfs}
\usepackage{epstopdf}
\usepackage{algorithm}
\usepackage{mathtools}
\usepackage[caption=false, font=scriptsize]{subfig}
\usepackage{algorithmic}
\usepackage{relsize}

\DeclareMathOperator*{\argmax}{arg\,max}

\newtheorem{definition}{Definition}

\newtheorem{theorem}{Theorem}

\newtheorem{remark}{Remark}

\definecolor{green}{rgb}{0, 0.5, 0}
\definecolor{pink}{rgb}{1, 0, 1}

\acrodef{agi}[AgI]{augmented information}
\acrodef{mec}[MEC]{mobile edge computing}
\acrodef{ldp}[LDP]{Lyapunov drift-plus-penalty}
\acrodef{lp}[LP]{linear programming}
\acrodef{cmdp}[CMDP]{constrained Markov decision process}
\acrodef{pdf}[pdf]{probability density function}
\acrodef{ue}[UE]{user equipment}
\acrodef{sfc}[SFC]{service function chain}
\acrodef{bp}[BP]{back-pressure}

\graphicspath{{figures/}}

\begin{document}

\title{Optimal Multicast Service Chain Control: Packet Processing, Routing, and Duplication}

\author{
	\IEEEauthorblockN{Yang Cai\IEEEauthorrefmark{1}, Jaime Llorca\IEEEauthorrefmark{2}, Antonia M. Tulino\IEEEauthorrefmark{2}\IEEEauthorrefmark{3}, Andreas F. Molisch\IEEEauthorrefmark{1}} 
	\IEEEauthorblockA{\IEEEauthorrefmark{1}University of Southern California, CA 90089, USA. Email: \{yangcai, molisch\}@usc.edu}
	\IEEEauthorblockA{\IEEEauthorrefmark{2}New York University, NY 10012, USA. Email: \{jllorca, atulino\}@nyu.edu}
	\IEEEauthorblockA{\IEEEauthorrefmark{3}University\`{a} degli Studi di Napoli Federico II, Naples 80138, Italy. Email: antoniamaria.tulino@unina.it}
	\thanks{
		An extended version of this paper is submitted to the IEEE Transactions on Communications \cite{cai2022multicast_arxiv}.
	}
}

\maketitle

\IEEEpubid{
\begin{minipage}{2\columnwidth}
	\centering
	{\footnotesize
	\vspace{80pt}
	\copyright\ 2021 {IEEE}. Personal use of this material is permitted. Permission from {IEEE} must be obtained for all other uses, in any current or future media, including reprinting/republishing this material for advertising or promotional purposes, creating new collective works, for resale or redistribution to servers or lists, or reuse of any copyrighted component of this work in other works.
	\cite{cai2021multicast}, DOI: \url{10.1109/ICC42927.2021.9500780}.
	}
\end{minipage}
}

\IEEEtitleabstractindextext{
\begin{abstract}
Distributed computing (cloud) networks, e.g., mobile edge computing (MEC), are playing an increasingly important role in the efficient hosting, running, and delivery of real-time stream-processing applications such as industrial automation, immersive video, and augmented reality. While such applications require timely processing of real-time streams that are simultaneously useful for multiple users/devices, existing technologies lack efficient mechanisms to handle their increasingly multicast nature, leading to unnecessary traffic redundancy and associated network congestion. In this paper, we address the design of distributed packet processing, routing, and duplication policies for optimal control of multicast stream-processing services. We present a characterization of the enlarged capacity region that results from efficient packet duplication, and design the first fully distributed {\em multicast traffic management} policy that stabilizes any input rate in the interior of the capacity region while minimizing overall operational cost. Numerical results demonstrate the effectiveness of the proposed policy to achieve throughput- and cost-optimal delivery of stream-processing services over distributed computing networks.
\end{abstract}}

\acresetall

\IEEEdisplaynontitleabstractindextext

\IEEEpeerreviewmaketitle


\section{Introduction}

\begin{figure}[t]
	\centering
	\subfloat[Multi-user conferencing.]{
		\includegraphics[width = .49 \linewidth]{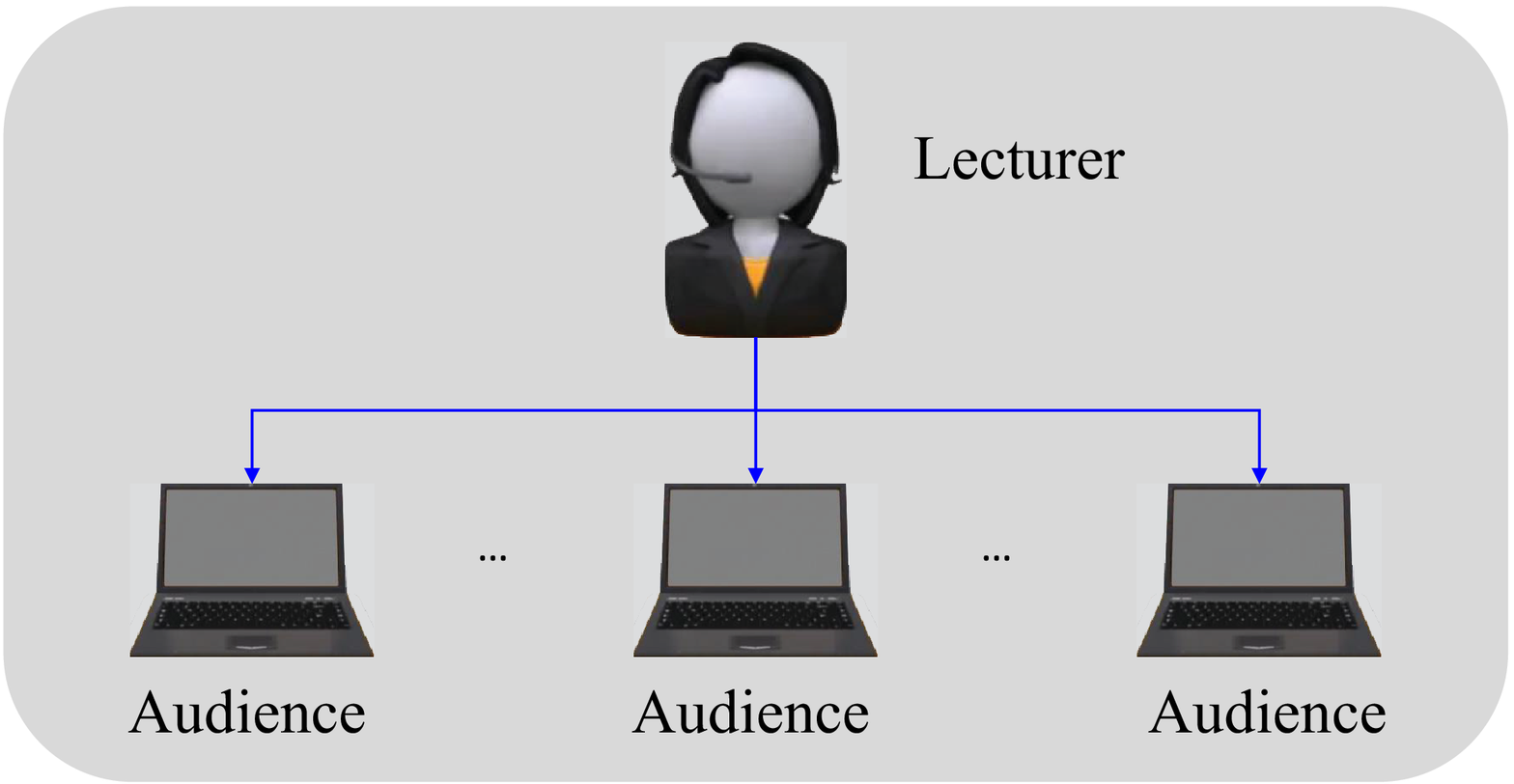}
		\label{fig:scene_1}
	}
	\subfloat[Smart vehicle system.]{
		\includegraphics[width = .49 \linewidth]{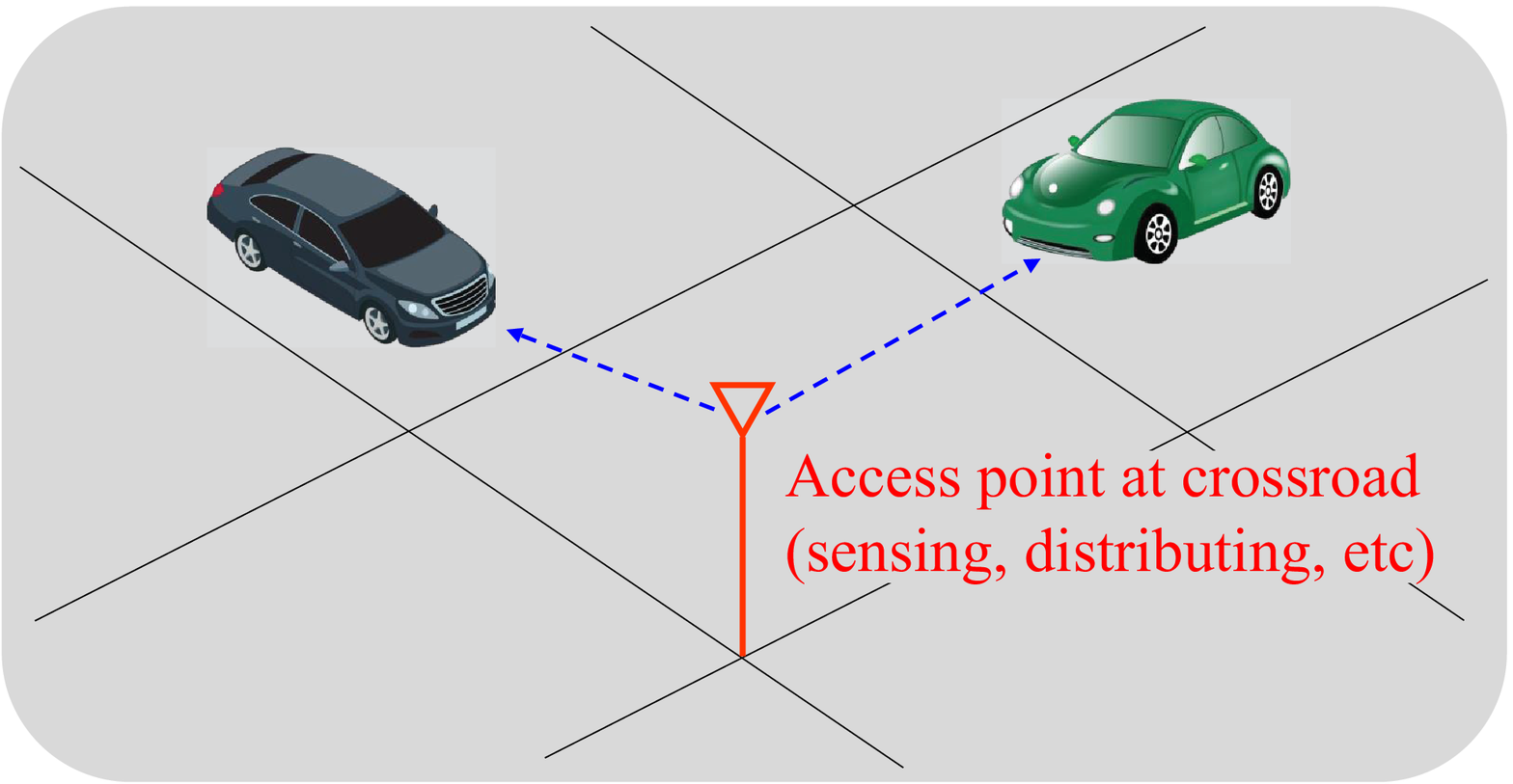}
		\label{fig:scene_2}
	}
	\caption{
		Two widely-used applications involving multicast network traffic: a) multiuser conferencing and b) vehicle coordination, which require the source information to be processed and delivered to multiple destination nodes.
	}
	\label{fig:V_time_1}
\end{figure}

The proliferation of real-time stream-processing applications such as augmented reality, telepresence, and industrial automation~\cite{cai2022metaverse,cai2022xpipelines_arxiv,cai2022CCC_arxiv}, is pushing the evolution of networking and cloud technologies in order to meet their stringent low latency and compute-intensive requirements \cite{Wel:B16}. Traditional approaches treat network and cloud resources separately, with fairly centralized core clouds handling the processing of compute-intensive tasks, while the network takes care of routing data streams from sources to the cloud, and back to their destinations. However, next-generation services can be decomposed into chains of individual functions that allows a more flexible and granular processing of data streams at distributed cloud locations. Service function chaining  precisely refers to the routing of traffic flows through an ordered sequence of service functions deployed at multiple cloud locations~\cite{Wel:B16}. %
In recent years, multicast data-streams (contents with multiple destinations) have become an increasingly dominant component of network traffic,%
\footnote{
	To clarify, the term {\em multicast} in this paper refers to content delivery to multiple destinations, not related to the wireless communication technique of transmitting data to multiple nodes simultaneously, as considered in \cite{FenLloTulMol:J18b}.
}
especially in the coming Internet of things (IoT) era. For example, multi-user conferencing (Fig. \ref{fig:scene_1}) requires to encode and deliver source information to several audiences; applications of another type, which can be summarized as joint decision making of multi-agent systems, including robot (or car) coordination in smart factory (or intelligent vehicle system, as shown in Fig. \ref{fig:scene_2}), also require the access point to distribute the sensing information/decided actions to multiple end nodes.

In order to maximize the benefit of distributed computing networks to support multicast services, two fundamental problems need to be addressed:
\begin{itemize}
	\item how to instantiate processing functions on edge/cloud servers and route the data-stream through them; 
	\item how to schedule and allocate network (computing and transmission) resources for different requests.
\end{itemize}

The first problem, usually referred to as \ac{sfc} optimization, involves jointly allocating tightly coupled cloud and network resources in order to decide where to run each service function and how to route service flows through the appropriate sequence of functions in order to maximize throughput and minimize overall operational cost. A number of recent works have addressed the \ac{sfc} optimization problem with the goal of either maximizing accepted service requests or minimizing overall resource cost~\cite{BarLloTulRam:C15,BarChoAhmBou:C15,BhaJaiSamErb:J16}. However, the problem is usually formulated under a static configuration, without taking into account increasingly prominent uncertain network conditions and time-varying service demands.

For the second problem, a closely related research field is {\em dynamic packet routing}, which has been extensively studied in the past, with two main celebrated mechanisms for decision making. On one hand, {\em source routing} schemes determine the entire route of the packet to the destination at the source node. In \cite{SinMod:J18}, a universal throughput-optimal source routing policy is designed for both unicast and multicast traffic. On the other hand, distributed routing schemes based on the {\em fluid model} determine packet routes based on local decisions on a hop-by-hop basis. The backpressure algorithm \cite{TasEph:J92} is an example of such policy that achieves throughput-optimal routing for unicast traffic. While, in general, source routing can achieve better delay performance, its centralized nature incurs additional overhead in collecting network-wide state information, making it more suitable for regimes with low congestion levels and relatively stable arrival rates. On contrary, distributed fluid-based algorithms only require local information exchange and decision making, and while they can suffer from inefficient {\em loopy} routes in low congestion scenarios, they are especially suitable for high congestion regimes. Besides, a recent study \cite{cai2022delay_arXiv,cai2021delay} proposes a distributed, backpressure-fashioned network control policy, which is designed to support services with stringent latency constraints.

Extensions of the above policies for \acp{sfc} have also been studied in recent works, either by introducing the computation flow \cite{FenLloTulMol:J18a} or constructing the layered graph \cite{ZhaSinLloTulMod:C18}. More concretely, \cite{ZhaSinLloTulMod:C18} investigates throughput-optimal service chain source routing for both unicast and multicast traffic; \cite{FenLloTulMol:J18a,cai2020mec} study throughput and cost optimal service chain distributed routing and resource allocation for unicast traffic (in particular, \cite{cai2020mec} addresses the related problems under a MEC network scenario). However, no throughput-optimal fully {\em distributed} policies have been designed for the multicast service chain control problem.

Motivated by the increasing multicast nature of next-generation real-time stream-processing services and scalating network congestion levels, in this paper, we focus on the design of throughput and cost optimal multicast service chain control policies. Multicast routing policies are of paramount importance to avoid excessive network congestion from unnecessary traffic redundancies. However, the main challenge in the design of distributed multicast routing policies is the difficulty to capture in-network packet duplication mechanisms that break flow conservation laws.

In this work, we provide the first formal analysis of fully distributed multicast routing policies (for arbitrary communication and computation services) that include joint packet processing, routing, and duplication. Our contributions can be summarized as follows: 
\begin{itemize}
  \item We characterize the enlarged multicast computing network capacity region that results when allowing in-network packet duplication. 
  \item We develop the first throughput- and cost-optimal fully distributed packet processing, routing, and duplication policy for multicast service chain control.
  \item We present numerical results demonstrating the enlarged multicast capacity region, and the tunable $[\mathcal{O}(V), \mathcal{O}(1/V)]$ cost-delay tradeoff associated with the proposed control policy.
\end{itemize}

\section{System Model}

\subsection{Cloud network}

We consider a wide-area distributed computing network, simply referred to as {\em cloud network}, modeled by graph $\Set{G} = (\Set{V}, \Set{E})$. Each node $i \in \Set{V}$ represents a network node with computing capabilities (e.g., core cloud, edge cloud, compute-enabled base station). Data can be transmitted from node $i$ to $j$ via network link $(i, j) \in \Set{E}$. We denote by $\delta_i^-$ and $\delta_i^+$ the incoming and outgoing neighbors of node $i$, respectively.

Assuming a time-slotted system, the available processing/ transmission resources, and associated costs, are defined as
\begin{itemize}
	\item $C_i$: the processing capacity, e.g., the number of computing cycles per time slot, at node $i$;
	\item $e_i$: the processing cost, i.e., the cost of running one unit of processing resource, at node $i$;
	\item $C_{ij}$: the transmission capacity, i.e., the data-stream size that can be transmitted in one time slot, on link $(i, j)$;
	\item $e_{ij}$: the transmission cost, i.e., the cost of transmitting one unit of data, on link $(i, j)$. 
\end{itemize}

\subsection{Service Chain}

The cloud network offers a set of services $\Phi$. Each service $\phi \in \Phi$ is modeled as an ordered chain of $(M_{\phi} - 1)$ functions, through which incoming packets must be processed. Functions can be executed at different network locations. While, for ease of exposition, we assume every cloud node can host any service function, it is straightforward to extend our model to limit the set of functions available at each cloud node. There are two parameters associated with each function: for the $m$-th function of service $\phi$, we define
\begin{itemize}
	\item $\xi_\phi^{(m)}$: the scaling factor, i.e., the output data-stream size per unit of input data-stream;
	\item $r_\phi^{(m)}$: the workload, i.e., the amount of computing resource required to process one unit of input data-stream.
\end{itemize}

We refer to the input and output data-streams of service $\phi$ as the stage $m$ and stage $m+1$ data-streams of service $\phi$, respectively. Data-streams are divided into packets of uniform length, and we assume that each packet can be processed separately.

In order to characterize the multicast nature of offered services, we assume each service $\phi\in\Phi$ is consumed by a set of destinations denoted by $\Set{D} = \{d_1, \cdots, d_D\}$ with $\Set{D} \subset \Set{V}$ and $|\Set{D}| = D$.

\subsection{Data Management}\label{sec:duplication}

In the unicast service control problem \cite{FenLloTulMol:J18a}, there are two relevant packet operations, i.e., processing and transmission. For multicast service control, we add the {\em packet duplication} operation to allow any network node to make two copies of any incoming packet.

Originally, prior to any duplication operation, each packet of a given service is associated with the entire destination set $\mathcal D$. After a duplication operation, each resulting copy is associated with a new destination set. The key requirement for any duplication operation is the {\bf coverage} of the original destination set, i.e., each destination node of the original packet must be present in the destination set of least one of the resulting copies. If the destination sets of the resulting copies do not overlap, the duplication operation is termed {\em efficient} (and {\em inefficient} otherwise).

To keep track of the changes in the destination sets after packet duplication operations, we introduce the concept of packet {\em duplication status}.

\begin{definition}[Duplication Status]
The duplication status of a packet, denoted by $q  = [q_1, \cdots, q_D]\in 2^{D}$, is a binary vector with $q_k = 1\ (k=1,\cdots, D)$ indicating that $d_k\in \Set{D}$ is one of its current destinations.
\end{definition}

In the above definition, $2^D$ is the set of indicator vectors corresponding to the power set of $\Set{D}$. In addition, we define a subset of it as $2^q \triangleq \{ s: s_k = q_k u_k\text{ with }u\in 2^D \}$, which collects $s$ whose entry must be $0$ if the entry is $0$ in $q$. Specially, $q = b_k$ (the binary vector with only the $k$-th entry equal to $1$) indicates a packet with only one destination $d_k$ (behaves as a unicast packet); and $q = \V{0}$ indicates a packet with no destination, which is not of interest and all the related quantities should be ignored.

We define a commodity as the collection of packets with the same $4$-tuple $(\phi, m, \Set{D}, q)$ description, i.e., service $\phi$, stage $m$, destination set $\Set{D}$, and duplication status $q$. To simplify the notation, we define $c\triangleq (\phi, m,\Set{D})$, and label a commodity as $(c, q)$.

Finally, we define the process of exogenous arrival of packets of commodity $(c, q)$ at node $i$ as $\big\{ a_i^{(c, q)}(t) : t\geq 0\big\}$. All the arriving processes are assumed to be i.i.d. over time slots and independent with each other, with mean rate $\mathbb{E}\big\{ a_i^{(c, q)}(t) \big\} = \lambda_i^{(c, q)}$, and finite second moment.

\section{Policy Space}\label{sec:policies}

In this section, we first present a general policy space for multicast service control, as well as the conditions for a policy to be admissible. We then describe an efficient policy space, by restricting the duplication process to be efficient, which does not reduce the performance (capacity region and the achievable optimal cost).

\subsection{General Policy Space}

We consider a general policy space for multicast service control, encompassing all packet processing, routing, and duplication policies. The decisions made by a policy in this space can be described by the following variables
\begin{align}
f(t) = \left\{ f_{i,\text{pr}}^{(c, q)}(t), f_{ij}^{(c, q)}(t) : \forall\, (c, q), i\in \Set{V}, (i,j) \in \Set{E} \right\}
\end{align}
which are the amount of packets of each commodity that are operated (processed or transmitted) on each interface.

A control policy is called {\em admissible}, if it makes decisions satisfying the following constraints:
1) non-negativity
\begin{align}\label{eq:cap_region_1}
f(t) \succeq 0\quad (\text{element-wise})
\end{align}
2) capacity constraints (recall that $c = (\phi, \Set{D}, m)$)
\begin{subequations}
\begin{align}
\tilde{f}_i(t) = \sum\nolimits_{(c, q)} r_{\phi}^{(m)} f_{i,\text{pr}}^{(c, q)}(t) \leq C_i,\quad\forall\,i\in \Set{V} \\
f_{ij}(t) =\sum\nolimits_{(c, q)} f_{ij}^{(c, q)}(t) \leq C_{ij},\quad\forall\,(i,j)\in \Set{E}
\end{align}
\end{subequations}
3) the generalized flow conservation and duplication law, for any commodity $c$ and $\forall\,k\in \{1,\cdots, D\}$:
\begin{align}\label{eq:flow_conserve}
\sum\nolimits_{\{ q: q_k = 1 \}} \big[ f_{\to i}^{(c, q)} + \lambda_i^{(c, q)} \big]
\leq \sum\nolimits_{\{ q: q_k = 1 \}} f_{i \to}^{(c, q)}
\end{align}
where $\{ q: q_k = 1 \}$ is the set of all the duplication status which indicates that $d_k \in \Set{D}$ is one of the current destinations; the incoming and outgoing flows are
\begin{subequations}
\begin{align}
f_{\to i}^{(c, q)} = \overline{ \Big\{ f_{\text{pr}, i}^{(c,q)}(t) + \sum\nolimits_{j\in \delta_i^-} f_{ji}^{(c,q)}(t) \Big\} } \\
f_{i \to}^{(c, q)} = \overline{ \Big\{ f_{i, \text{pr}}^{(c,q)}(t) + \sum\nolimits_{j\in \delta_i^+} f_{ij}^{(c,q)}(t) \Big\} }
\end{align}
\end{subequations}
with the processed flow $f_{\text{pr}, i}^{(c,q)}(t) = f_{\text{pr}, i}^{(\phi, m, \Set{D}, q)}(t)$ defined as
\begin{align}\label{eq:cap_region_2}
f_{\text{pr}, i}^{(\phi, m, \Set{D}, q)}(t)
= \begin{cases} 0 & m = 1 \\
\xi_\phi^{(m-1)} f_{i, \text{pr}}^{(\phi, m-1, \Set{D}, q)}(t) & m > 1 \end{cases}
\end{align}
and $\avg{ \cdot }$ denotes the long-term average operator
\begin{align}
\avg{z(t)} \triangleq \lim_{T\to\infty} \frac{1}{T} \sum\nolimits_{t=1}^{T}{ z(t) }.
\end{align}
The generalized flow conservation and packet duplication law \eqref{eq:flow_conserve} holds because of the {\bf coverage} requirement (see previous section). For any destination $d_k$ of an incoming packet, there is {\em at least} one outgoing packet (one of its copies if duplicated, or itself otherwise) with $d_k$ in its destination set.

The instantaneous overall resource cost incurred by the above policy is defined as
\begin{align}
h(t) = \sum\nolimits_{i\in \Set{V}} e_i \tilde{f}_i(t) + \sum\nolimits_{(i, j)\in \Set{E}} e_{ij} f_{ij}(t)
\end{align}
and its long-term average $\avg{ h(t) }$ is employed to characterize the cost performance of the policy. Furthermore, we denote by $h^\star(\V{\lambda})$ the optimal cost that can be achieved by the general policy space, under the arrival rate $\V{\lambda}$.

Finally, we define the {\em capacity region} $\Lambda$ of the cloud network as the set of all arrival vectors $\V{\lambda} = \big\{ \lambda_i^{(c, q)} \big\}$, such that there exists a control policy satisfying \eqref{eq:cap_region_1} -- \eqref{eq:cap_region_2}.

\subsection{Efficient Policy Space}

We now define an efficient policy space as a subset of the general space, by requiring all the duplication operations to be efficient. More concretely, if two copies are created from a packet by a duplication operation, then
\begin{align}\label{eq:efficient_duplication}
q = s + r
\end{align}
with $q, s, r \in 2^D$ denoting the duplication status of the original packet and the two copies, respectively.

When a duplication is performed in an efficient way, for any destination node of a particular incoming packet, there will be {\em exactly} one outgoing packet steering to it. In this case, the flow conservation and duplication law can be cast as 
\begin{align}\label{eq:flow_conserve_eq}
\sum\nolimits_{\{ q: q_k = 1 \}} \big[ f_{\to i}^{(c, q)} + \lambda_i^{(c, q)} \big]
= \sum\nolimits_{\{ q: q_k = 1 \}} f_{i \to}^{(c, q)}.
\end{align}

By restricting to the efficient space, we eliminate repeated delivery of identical content to the same destination node, which is beneficial for 1) alleviating the network traffic, as well as 2) reducing the resource cost. Specially, this is true when comparing with the optimal policy of the general space. As a consequence, the efficient policy space can achieve the same {\em capacity region} as the general space,
and the achievable {\em optimal cost} by the efficient policy space equals to $h^\star(\V{\lambda})$.

\section{Queueing System}

We construct the queueing system by creating a queue $Q_i^{(c, q)}(t)$ for each commodity $(c, q)$ at each node $i$.

The efficient policy space is considered, and we describe a typical operation procedure for a packet in one time slot in the following. Suppose a packet of duplication status $q$ is selected for operation (processing or transmission) on a certain interface, we need to decide whether it will be duplicated or not.%
\footnote{
	We consider the scheme where each packet is duplicated at most once in a time slot. Compared to a more general scheme without this restriction, the considered scheme just splits duplications into multiple steps, and that does not increase traffic, while only increasing delay by a finite amount of slots, which does not affect the capacity region or the cost performance.
}
If a packet is duplicated, only one copy is {\em operated} on the interface, while the other copy is {\em reloaded} to the queueing system at the end of the time slot (i.e., it is not involved in any other decisions in the current time slot).

The above description motivates us to involve the {\em posterior} duplication status $s\in 2^q$ in the formulation, which is the status of the operated copy (and by \eqref{eq:efficient_duplication}, the status of the reloaded copy is $q-s$). Specially, the case $q = s$ indicates that the packet is not duplicated. To sum up, the $(q, s)$-pair specifies a duplication decision.

\begin{figure}[t]
\centering
\includegraphics[width = .9 \columnwidth]{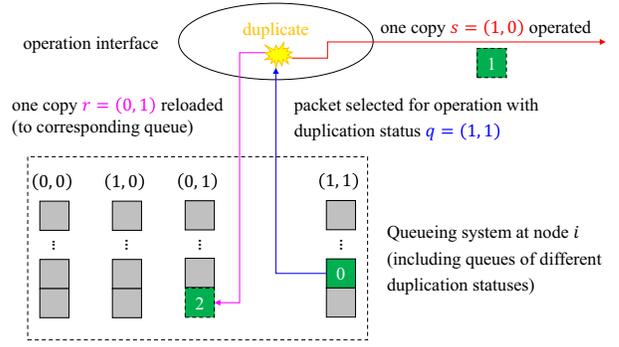}
\caption{
	Structure of the established queueing system at any network node $i$ (for an application with $D = 2$ destination nodes). In order to distinguish packets with different current destination set, we create $2^D = 4$ queues corresponding to the $4$ duplication statuses, i.e., $\{0, 1\}^2$. In each time slot, in addition to the scheduling decision, i.e., which packets will be operated at which interface (the blue link), we also need to make a duplication decision, i.e., whether to split the network flow or not, and how (mathematically, to determine the duplication status of the operated copy $s$ and the reloaded copy $r$). When $r = (0, 0)$ (and thus $s = q$), the packet is operated without changing the assigned destination set, and no copy is created in this case (in fact, node $i$ does not need to manage packets in the $(0, 0)$ queue; we present it in the figure just for completeness).
}
\end{figure}

\subsection{Queueing Dynamics}\label{sec:queueing_system}

Let $x_{i,\text{pr}}^{(c, q, s)}(t)$ and $x_{ij}^{(c, q, s)}(t)\ (j\in\delta_i^+)$ be the amount of packets of commodity $(c, q)$ desired by the output interfaces, on which the duplication decision $(q, s)$ will be performed. In general, the queueing dynamics is given by
\begin{align}\label{eq:q_dynamic}\begin{split}
Q_i^{(c, q)}(t+1) & \leq \Big[  Q_i^{(c, q)}(t) - \sum\nolimits_{s\in 2^q} \mu^{(c, q, s)}_{i\to}(t) \Big]^+ \\
& \quad + \mu^{(c, q)}_{\to i}(t) + a_i^{(c, q)}(t)
\end{split}\end{align}
where the outgoing flow is
\begin{align}
\mu^{(c, q, s)}_{i\to}(t) = x_{i,\text{pr}}^{(c, q, s)}(t) + \sum\nolimits_{j\in \delta_i^+}{  x_{ij}^{(c, q, s)}(t) }
\end{align}
and the (controllable) incoming flow is
\begin{align}\label{eq:income}
\mu^{(c, q)}_{\to i}(t) & = \sum\nolimits_{s\in 2^{\bar{q}}} \Big[ x_{\text{pr},i}^{(c, q+s, q)}(t) + \sum\nolimits_{j\in \delta_i^-}{ x_{ji}^{(c, q+s, q)}(t) } \Big] \nonumber \\
& \quad + \sum\nolimits_{s\in 2^{\bar{q}}} \mu^{(c, q+s, s)}_{i\to}(t)
\end{align}
with $\bar{q} = \V{1} - q$; and $[z]^+ \triangleq \max\{z, 0\}$. The two lines in \eqref{eq:income} represent the operated and the reloaded packets, respectively. The reloaded part is explained as follows: a packet of status $q+s$ is duplicated, with the copy of status $s$ operated; thus the other copy of status $(q+s) - s = q$ will be reloaded.

Specially, \eqref{eq:q_dynamic} does not apply to queues of {\em destination state}, i.e., $Q_{i_0}^{(c_0, q)}(t)$ with $i_0 = d_k \in \Set{D}$ and $c_0 = (\phi, M_{\phi}, \Set{D})$. If a packet of commodity $(c_0, q)$ (with $q_k = 1$) arrives at $i_0$, it will be consumed. But due to the multicast nature of the packet (in general), it will be duplicated into two copies of status $b_k$ and $q' = q-b_k$ (and thus $q'_k = 0$), with the copy of $b_k$ departing the network, and the other copy reloaded to the queue $q'$. Therefore, the queue $q$ is always empty, while queue $q'$ receives an extra packet compared to the general case. To sum up, in this case, the queueing dynamics is given by

\vspace{-.15in}
\begin{small}
\begin{align}\label{eq:q_dynamic_des}
Q_{i}^{(c, q)}(t+1) \leq \begin{cases}
0 & q_k = 1 \\
R + \mu^{(c, q+b_k)}_{\to i}(t) + a_i^{(c, q+b_k)}(t)  & q_k = 0
\end{cases}
\end{align}
\end{small}
\hspace{-.1cm}where $R$ is the right-hand-side of \eqref{eq:q_dynamic}.

\subsection{Problem Formulation}

Based on the queueing system introduced in the previous section, mathematically, the multicast service chain control problem is formulated as
\begin{subequations}\label{eq:formulation}
\begin{align}
& \min_{\V{x}(t)} \quad \avg{ \E{h(t)} } \label{eq:obj_func} \\ 
& \operatorname{s.t.} \quad \text{stabilizing the queueing system \eqref{eq:q_dynamic} -- \eqref{eq:q_dynamic_des}} \\
& \hspace{.4 in} x_{\text{pr},i}^{(\phi, m+1, \Set{D}, q, s)}(t) = \xi_{\phi}^{(m)} x_{i,\text{pr}}^{(\phi, m, \Set{D}, q, s)}(t) \\
& \hspace{.4 in} \tilde{x}_i(t) \triangleq \sum_{(c, q, s)} r_{\phi}^{(m)} x_{i,\text{pr}}^{(c, q, s)}(t) \leq C_i\quad\forall\,i\in \Set{V} \label{eq:cap_con_1}\\
& \hspace{.4 in} x_{ij}(t) \triangleq \sum_{(c, q, s)} x_{ij}^{(c, q, s)}(t) \leq C_{ij}\quad\forall\,(i,j)\in \Set{E} \label{eq:cap_con_2}\\
& \hspace{.4 in} \V{x}(t)\succeq 0\ (\text{element-wise}). \label{eq:geq0}
\end{align}
\end{subequations}

\begin{remark}
In the above formulation, note that decisions $\V{x}(t)$ are made regardless of the available packets in the queue, it can happen that the requests raised by the interfaces cannot be satisfied. In that case, {\em dummy} packets will be created and sent to the interface to compensate for the lack of actual packets, as is considered in \cite{FenLloTulMol:J18a} for the unicast case.
\end{remark}

\section{Capacity Region}

In this section, we present a characterization for the capacity region of cloud network with multicast flows, which is based on the celebrated {\bf fact} \cite{Nee:B10} that there exists a stationary randomized policy $*$ to stabilize any point within the capacity region, while achieving the optimal objective (cost) value.

\begin{theorem}
An arrival vector $\V{\lambda}$ is within $\Lambda$ if and only if
there exists flow variables $\V{f} = \big\{ f_{i,\text{pr}}^{(c,q,s)}, f_{ij}^{(c,q,s)} \big\} \succeq 0$
together with probability values $\{ \beta_{i}^{(c, q, s)} \}_{(c,q,s)}$ and $\{ \beta_{ij}^{(c, q, s)} \}_{(c,q,s)}$ for $\forall\, i\in \Set{V}$, $(i,j) \in \Set{E}$ such that
\begin{small}
\begin{subequations}\begin{align}
& \sum_{s\in 2^{\bar{q}} } \Big[ f_{\text{pr},i}^{(c, q+s, q)} + \sum_{j\in \delta^-_i}{ f_{ji}^{(c, q+s, q)} } + f_{i, \text{pr}}^{(c, q+s, s)} + \sum_{j\in \delta^+_i}{ f_{ij}^{(c, q+s, s)} } \Big] \nonumber \\
& \hspace{.5in} + \lambda^{(c, q)}_{i}  \leq \sum\nolimits_{s\in 2^q} \Big[ f_{i,\text{pr}}^{(c, q, s)} + \sum\nolimits_{j\in \delta^+_i}{ f_{ij}^{(c, q, s)} } \Big] \\
& \hspace{.2in} f_{\text{pr},i}^{(\phi, m+1, \Set{D}, q, s)} = \xi_\phi^{(m)} f_{i,\text{pr}}^{(\phi, m, \Set{D}, q, s)} \\
& \hspace{.6in} f^{(c, q, s)}_{i,\text{pr}} \leq \big( C_i/r_\phi^{(m)} \big) \beta^{(c, q, s)}_{i} \label{eq:ncr_2} \\
& \hspace{.6in} f^{(c, q, s)}_{ij} \leq \beta^{(c, q, s)}_{ij} C_{ij}.
\end{align}\end{subequations}
\end{small}
\hspace{-.2 cm}and the stationary randomized policy $*$ specified by the probability values $\beta$ makes decisions $\V{x}^*(t)$ such that
\begin{align}
\avg{ \E{h( \V{x}^*(t) )} } = h^\star(\V{\lambda})
\end{align}
with $h^\star(\V{\lambda})$ denoting the optimal cost that can be achieved when the arrival vector is $\V{\lambda}$.
\end{theorem}

\begin{IEEEproof}
The result is derived by applying the {\bf fact} to the queueing system in Section \ref{sec:queueing_system} \cite{Nee:B10}. Details can be found in \cite{cai2022multicast_arxiv}.

The policy $*$ is defined as follows. For each interface, select the commodity $(c, q)$ and the duplication action $(q, s)$ independently in every time slot according to the probability value $\beta$; duplicate the packets according to $(q, s)$, and use all the available resource to operate the copies of status $s$.
\end{IEEEproof}

\section{Control Policy Design}\label{sec:control_policy}

Problem \eqref{eq:formulation} can be solved by Lyapunov drift-plus-penalty (LDP) approach \cite{Nee:B10}, as is shown in the following section.

\subsection{The LDP Approach}

We first define the Lyapunov function as $L(t) = \|\V{Q}(t)\|_2^2/2$ with $\V{Q}(t) = \big\{ Q_i^{(c,q,s)}(t) \big\}$, quantifying the current network congestion, and define the drift as $\Delta(t) = L(t+1) - L(t)$.

The LDP approach advocates to minimize (the upper bound of) a linear combination of the Lyapunov drift $\Delta(t)$ and the objective function $h(t) = h( \V{x}(t) )$ weighted by a tunable parameter $V$, given by \cite{Nee:B10}
\begin{align}\begin{split}\label{eq:dpp_bound}
\Delta(t) + V h(t) \leq B - \sum_{i\in \Set{V}} \sum_{(c, q, s)} w_{i}^{(c, q, s)} x_{i, \text{pr}}^{(c, q, s)}(t) \\
- \sum_{(i, j)\in \Set{E}} \sum_{(c, q, s)} w_{ij}^{(c, q, s)} x_{ij}^{(c, q, s)}(t)
\end{split}\end{align}
where $B$ is a constant, and the weights are given by
{\small \begin{subequations} \label{eq:weight} \begin{align}
w_{i}^{(c,q,s)} & = \frac{ Q_i^{(c,q)}(t) - Q_i^{(c,q-s)}(t) - \xi_\phi^{(m)} Q_i^{(c',s)}(t) }{ r_\phi^{(m)} } - V  e_{i} \label{eq:weight1}\\
w_{ij}^{(c,q,s)} & = Q_i^{(c,q)}(t) - Q_i^{(c,q-s)}(t) - Q_j^{(c,s)}(t) - V e_{ij} \label{eq:weight2}
\end{align}\end{subequations}}
\hspace{-.1cm}where $c' = (\phi, m+1, \Set{D})$.

The constraints on the decision variables $\V{x}(t)$ are given by \eqref{eq:cap_con_1}, \eqref{eq:cap_con_2} and \eqref{eq:geq0}, which leads to a solution in the form of {\em max-weight}, presented in the following section.

\subsection{Control Policy}\label{sec:control_alg}

Note that minimizing \eqref{eq:dpp_bound} can be completed separately on each interface (due to the additive form). The processing (or transmission) decisions are made by the following steps: for each node $i\in \Set{V}$ (or each link $(i, j)\in \Set{E}$),

1) calculate the weight for each tuple $(c,q,s)$ according to \eqref{eq:weight1} (or \eqref{eq:weight2}), based on the observed queue status;

2) find the tuple $(q, s, c)$ with the largest weight, i.e.,
\begin{align}
(q, s, c)^\star = \argmax_{(q,s,c)} \ w_i^{(q, s, c)}\ \big(\text{or } w_{ij}^{(q, s, c)} \big);
\end{align}

3) the optimal flow assignment is given by
{\small \begin{align}
x_{i,\text{pr}}^{(q, s, c)}(t) & = \frac{C_i}{ r_{\phi^\star}^{(m^\star)} } \,
\mathbb{I}\left\{ (q, s, c) == (q, s, c)^\star, w_{ij}^{(q, s, c)^\star}(t) > 0 \right\} \nonumber \\
x_{ij}^{(q, s, c)}(t) & = C_{ij} \, \mathbb{I}\left\{ (q, s, c) == (q, s, c)^\star, w_{ij}^{(q, s, c)^\star}(t) > 0 \right\}
\end{align}}
\hspace{-0.15cm}where $\mathbb{I}\{\cdot\}$ denotes the indicator function, which equals to $1$ only when the two conditions are both satisfied.

The developed algorithm only requires local information exchange and decision making, which can be implemented in a fully distributed manner.

\subsection{Performance Analysis}
We evaluate the performance of the proposed algorithm in the following theorem, using the achievable optimal cost as the benchmark.

\begin{theorem}
For any arrival vector $\V{\lambda}$ that is in the interior of the capacity region, the queue backlog and the cost achieved by the proposed algorithm satisfy
\begin{align}
\avg{ \E{\| \V{Q}(t) \|_1} } & \leq \frac{B}{\epsilon} + \left[ \frac{h^\star(\V{\lambda}+\epsilon\V{1}) - h^\star(\V{\lambda}) }{ \epsilon } \right] V \label{eq:delay_performance} \\
\avg{ \E{h(t)} } & \leq h^\star(\V{\lambda}) + \frac{B}{V} \label{eq:cost_performance} 
\end{align}
for any $\epsilon > 0$ such that $\V{\lambda}+\epsilon\V{1}\in \Lambda$.
\end{theorem}

\begin{IEEEproof}
The proof closely follows the philosophy of the proof of Theorem 2 in \cite{FenLloTulMol:J18a}.
\end{IEEEproof}

The above theorem reveals the $[\mathcal{O}(V), \mathcal{O}(1/V)]$ tradeoff between the delay (which is proportional to queue backlog by Little's theorem) and cost performance achieved by the proposed algorithm. In addition, for any fixed $V$, the queue backlog is mean rate state (i.e., $\avg{ \E{ \| \V{Q}(t) \|_1 } } < \infty$), implying that the proposed algorithm is throughput-optimal.

\subsection{Complexity Issue}

Finally, we analyze the complexity of the proposed algorithm, from both the communication and computation aspects.

\subsubsection{Communication Overhead}

The proposed algorithm requires local exchange of queue backlog information in every time slot. In contrast to transmitting the entire queueing status $\sim \mathcal{O}(2^D)$ in every time slot, we take advantage of the underlying max-weight structure of the proposed algorithm. More concretely, in every time slot, the proposed algorithm selects one commodity to operate on each interface; as a result, only one element of the queueing vector of node $j$ changes. Therefore, the number of queues with varying backlogs is $\sim \mathcal{O}(\delta_{\max}^+)$, where $\delta_{\max}^+$ is the largest incoming degree. By transmitting information related to only these queues, the communication overhead can be greatly reduced.

\begin{figure}[t]
	\centering
	\includegraphics[width = .6 \columnwidth]{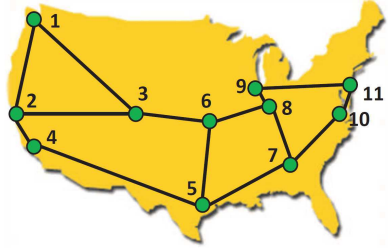}
	\caption{The continental US Abilene network.}
	\label{fig:network}
\end{figure}

\subsubsection{Computational Complexity} \label{sec:complexity}

In every time slot, each node needs to calculate the weights of all $(c, q, s)$ tuples in order to decide the best commodity to operate on, and make the duplication decision. It can be shown that for a fixed content $c$, the number of possible $(q, s)$ pairs is $3^D - 2^D \sim \mathcal{O}(3^D)$. Although to calculate the weight for each $(c, q, s)$ (at each interface) by \eqref{eq:weight} requires only simple algebraic operations, the number of the tuples grows exponentially with the size of the destination set, and there is no quick way to reduce the computation complexity of the algorithm to polynomial-time.\footnote{This is determined by the combinatorial nature of the multicast problem. Another solution to the multicast problem provided by \cite{ZhaSinLloTulMod:C18} requires to solve the minimum Steiner tree problem to determine the route for each packet, which is a NP-complete problem.}

To sum up, with more destination nodes, we can envision larger performance improvement compared to the {\bf simple approach} that treats them as individual unicast flows (since the proposed method has the potential to reuse more intermediate results). However, the algorithm also becomes more computationally demanding, making it not suitable to apply to large scale networks. Developing an efficient, approximate algorithm is the topic of our ongoing research work, and a polynomial-time heuristic algorithm will be reported in \cite{cai2022multicast_arxiv}.

\section{Numerical Results}

We perform the numerical experiments based on the continental US Abilene network, as is shown in Fig. \ref{fig:network}. The processing capability of each node is $C_i = 20$ CPUs, and the processing cost is $e_i = 0.5\ /$CPU per second. The cloud network links exhibit homogeneous transmission capabilities and costs, given by $C_{ij} = 10$ Gbps, and $e_{ij} = 1\ /$Gb. We set the length of each time slot as $\tau = 1$ ms, and unify the size of each packet as $F = 1\ \text{kb}$.

Two services are provided by the cloud network, each consisting of $2$ functions, with the following parameters
\begin{align}
\phi_1 &:\ \xi_1^{(1)} = 1,\ \xi_1^{(2)} = 2;\ 1/r_1^{(1)} = 300,\ 1/r_1^{(2)} = 400 \nonumber \\
\phi_2 &:\ \xi_2^{(1)} = \frac{1}{3},\ \xi_2^{(2)} = \frac{1}{2};\ 1/r_2^{(1)} = 200,\ 1/r_2^{(2)} = 100 \nonumber
\end{align}
where $1/r_{\phi}^{(m)}\ [\text{Mbps}/\text{CPU}]$ denotes the supportable input size given $1$ CPU resource.

We consider any destination set $\Set{D}$ consisting of two nodes selected from $\{7, 8, 9, 10, 11\}$ (e.g., $\{7, 10\}$), and hence there are $10$ possible destination sets in total. Each destination set can request both services $\phi_1$ and $\phi_2$, which originate from any source node in $\{1,2, 3, 4\}$. The packets of commodity $(c, q) = (\phi_i, 1, \Set{D}, \V{1})$ ($i =1, 2$) arrive at each source node, and it is modeled by i.i.d. Poisson process, independent of each other, with parameter $\lambda$.

We employ the {\bf simple approach} (see Section \ref{sec:complexity}) as the baseline for comparison, i.e., treating data-streams for different destination nodes as separated unicast flows. A more comprehensive comparison of the proposed approach with existing multicast techniques will be reported in \cite{cai2022multicast_arxiv}.

\subsection{Capacity Region}

\begin{figure}[t]
	\centering
	\includegraphics[width = .4\textwidth]{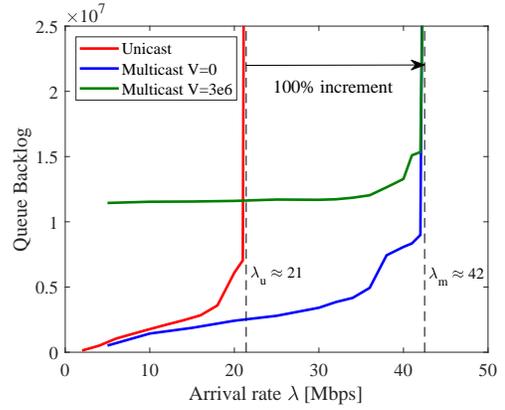}
	\caption{The capacity region achieved by the multicast algorithms (with $V = 0$ and $V = 3\times 10^6$), as well as the unicast-based solution.}
	\label{fig:cap_region}
\end{figure}

We first study the capacity region of the cloud network with multicast flow (using the proposed algorithm), compared with the achieved capacity region by treating the problem as separate unicast problems. The initial queue backlog is set as $\V{Q}(0) = \V{0}$, and we observe the system for $10^6$ time slots. The stable queue backlogs are recorded under various $\lambda$ values. If the queue keeps growing at the end of the period, the stable queue length is set as $\infty$.

The results is shown in Fig. \ref{fig:cap_region}. It is obvious that the queue backlog grows monotonously with the arrival rate for all the three cases. Then we focus on the queue backlog performance of the proposed algorithm under various values of $V$. We find that a larger value of $V$ results in a heavier queue backlog; however, the two values $V = 0$ and $V = 3\times 10^6$ lead to an identical critical point $\lambda_{\text{m}} \approx 42$ Mbps, which can be interpreted as the boundary of the capacity region. The result validates the conclusion that the proposed algorithm, using any fixed value of $V$, always achieves finite queue backlog within the capacity region, and therefore is throughput-optimal. Finally, we compare the capacity regions achieved by the proposed algorithm with the unicast-based solution, which is $\lambda_\text{u} \approx 21$ Mbps. An increment of $100\%$ is gained, by making smart duplication decision, which reuses some intermediate results to fully exploit the available resource.

\subsection{Delay-Cost Tradeoff}

\begin{figure}[t]
	\centering
	\includegraphics[width = .4\textwidth]{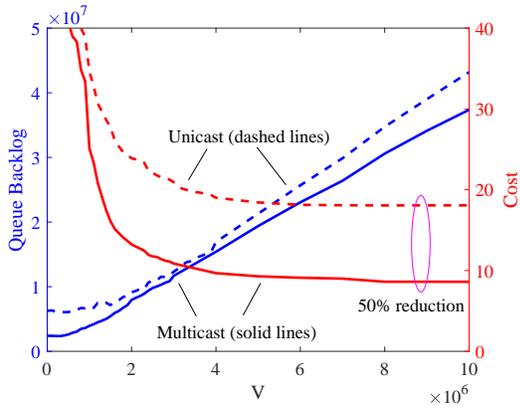}
	\caption{The queue backlog and cost performance of the proposed algorithm under various values of $V$.}
	\label{fig:tradeoff}
\end{figure}

Next, we study the queue backlog, as well as the cost performance of the proposed algorithm under various $V$. The arrival rate is selected as $\lambda = 20$ Mbps. The results are compared with the unicast-based solution.

The results are depicted in Fig. \ref{fig:tradeoff}. Visually, it exhibits a $[\mathcal{O}(V), \mathcal{O}(1/V)]$ tradeoff between the queue backlog and the resource cost, as is established in \eqref{eq:delay_performance} and \eqref{eq:cost_performance}. Considering the decreasing rate, we anticipate the optimal cost of the proposed algorithm to be $8$, which reduces by $50\%$ when comparing with the optimal cost $17$ achieved by the unicast-based solution. Again, the reduction is thanks to the reuse gain as is explained in the previous experiment. A larger gain can be expected for a destination set with more nodes, but this comes at the price of increasing the algorithm complexity.

\section{Conclusions}

In this paper, we investigated the problem of cloud network control in the presence of multicast flows. We proposed a queueing system that allows flow-level (rather than packet-wise) decision making, and presented an efficient policy space that is cost-optimal. The characterization of the new capacity region was presented, and we developed a fully distributed control algorithm guided by Lyapunov optimization theory. Numerical results showed the performance gain of the proposed algorithm over the unicast-based solution, in terms of the capacity region and the achieved resource cost.

\ifCLASSOPTIONcaptionsoff
  \newpage
\fi

\end{document}